\documentclass[aps,prb,twocolumn,superscriptaddress,longbibliography]{revtex4-1}

\usepackage[caption=false]{subfig}
\usepackage[colorlinks=true,citecolor=blue]{hyperref} 
\usepackage{graphicx}
\usepackage{physics}
\usepackage{bm}
\usepackage{amsmath}
\usepackage{comment}
\usepackage{amssymb}
\usepackage{qcircuit} 
\usepackage{blindtext}
\DeclareGraphicsExtensions{.png,.jpg,.eps}
\usepackage{xcolor}

\newcommand{\XW}[1] {\color{purple}XW: {#1}\color{black}\normalsize}

\begin{document}

\author{Marcel Niedermeier}
\affiliation{Department of Applied Physics, Aalto University, 02150 Espoo, Finland}
\affiliation{Univ. Grenoble Alpes, CEA, Grenoble INP, IRIG, Pheliqs, 38000 Grenoble, France}

\author{Adrien Moulinas}
\affiliation{Univ. Grenoble Alpes, CEA, Grenoble INP, IRIG, Pheliqs, 38000 Grenoble, France}

\author{Thibaud Louvet}
\affiliation{Univ. Grenoble Alpes, CEA, Grenoble INP, IRIG, Pheliqs, 38000 Grenoble, France}

\author{Jose L. Lado}
\affiliation{Department of Applied Physics, Aalto University, 02150 Espoo, Finland}

\author{Xavier Waintal}
\affiliation{Univ. Grenoble Alpes, CEA, Grenoble INP, IRIG, Pheliqs, 38000 Grenoble, France}

\title{Solving the Gross-Pitaevskii equation on multiple different scales using the quantics tensor train representation}

\begin{abstract}
Solving partial differential equations of highly featured problems represents a formidable challenge, where reaching high precision across multiple length scales can require a prohibitive amount of computer memory
or computing time. However, the solutions to physics problems typically have structures operating on different length scales, and as a result exhibit a high degree of compressibility. Here, we use the quantics tensor train representation to build a solver for the time-dependent Gross-Pitaevskii equation. 
We demonstrate that the quantics approach generalizes well to the presence of the non-linear term in the equation. We show that we can resolve phenomena across length scales separated by seven orders of magnitude in one dimension within one hour on a single core in a laptop, greatly surpassing the capabilities of more naive methods. We illustrate our methodology with various modulated optical trap potentials presenting features at vastly different length scales,
including solutions to the Gross-Pitaevskii equation on two-dimensional grids above a trillion points ($2^{20} \times 2^{20}$). This quantum-inspired methodology can be readily extended to other partial differential equations combining spatial and temporal evolutions, providing a powerful method to solve highly featured differential equations at unprecedented length scales.
\end{abstract}

\date{\today}

\maketitle

\section{Introduction}

The development of quantum computing has had two unexpected side effects. First, it is being realized that many mundane mathematical problems could be put in the framework of the quantum many-body (many qubits) 
formalism~\cite{Arrazola2020,PhysRevA.96.022329,KukHyunHan2002,Narayanan,Tang2019,Rnnow2014}. Second, one finds that apparently exponentially hard problems may not all be as hard as initially thought, after all~\cite{PhysRevX.10.041038,PRXQuantum.5.010308,PhysRevLett.129.090502,PhysRevLett.128.030501,QFT_small_entanglement}. In fact, some of these problems may not even be hard classically as demonstrated by the successes of tensor network methods~\cite{PhysRevLett.69.2863,RevModPhys.77.259,Ors2014,Jiang2019,Xu2024,PhysRevX.10.021042,Yan2011}.
A problem that can potentially benefit from a quantum-inspired speed-up is the solution of partial differential equations~\cite{Zanger2021} of the form $\partial_t f(t, \boldsymbol{r}) = F(t, \boldsymbol{r}, f)$.
Here $f(t=0, \boldsymbol{r})$ is the initial condition and $F$ is a possibly non-linear function of $f$ and its derivatives. Such problems are ubiquitous in the physical sciences, including the Fokker-Planck~\cite{Fokker1914,planck1917satz,PhysRev.162.186,Jordan1998},  Schrödinger and Kohn-Sham
equations~\cite{Gross1961,pitaevskii1961vortex,PhysRevLett.52.997,PhysRevLett.82.3863}, electromagnetism\cite{KaneYee1966,monk2003finite,Teixeira2023}, 
computational fluid dynamics~\cite{Harlow1965,Issa1986,ferziger2019computational},
and micromagnetism~\cite{landau1935theory,Gilbert2004,Fidler2000}, among others. Except for a few special cases, these problems are not analytically solvable and require a numerical solution.

\begin{figure}[t!]
    \centering
    \includegraphics[width=\columnwidth]{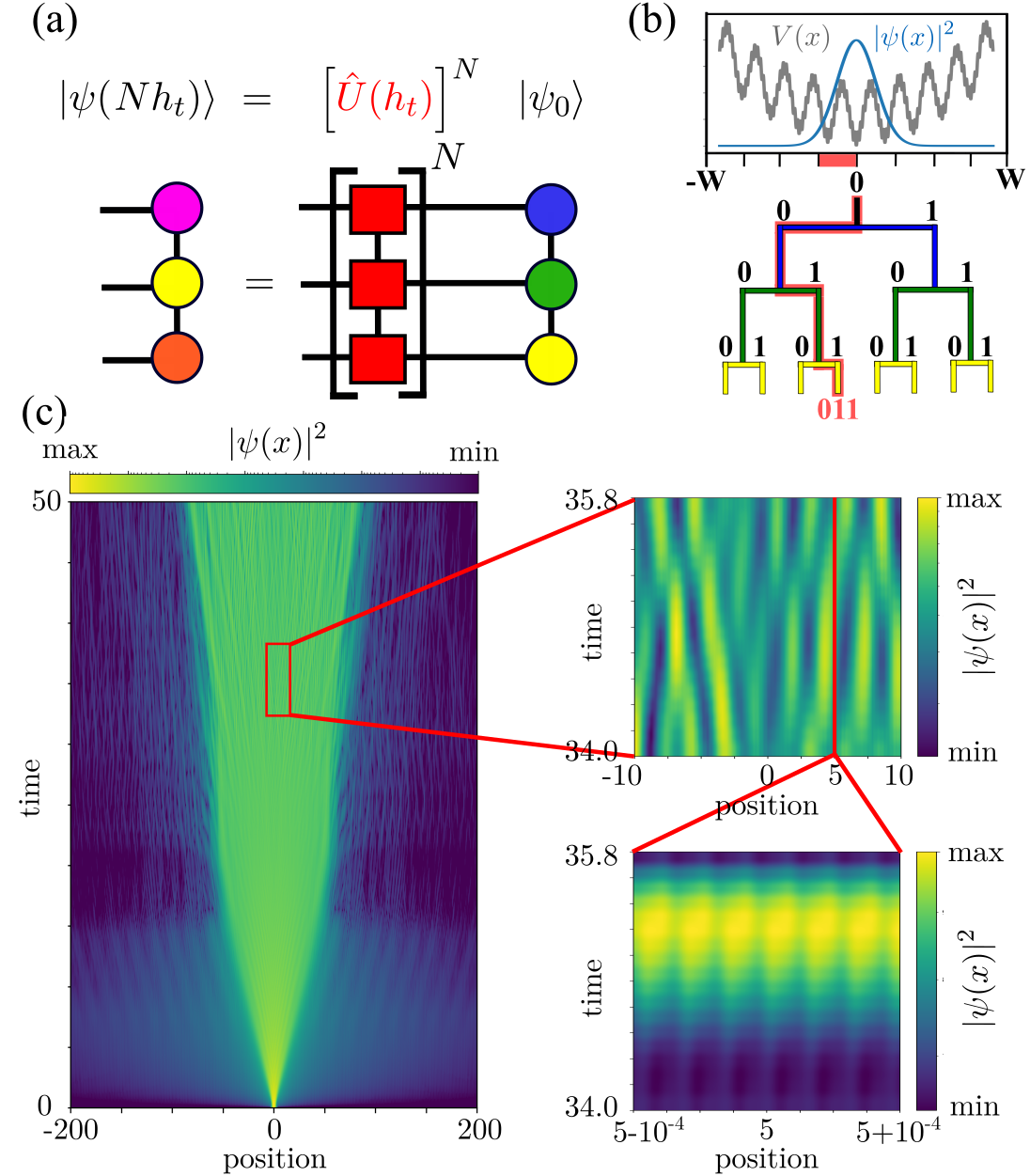}
    \caption{Overview.  (a) We time-evolve an initial state $\ket{\psi_0}$ in $N$ time slices. Each incremental time-evolution operator $\hat U(h_t)$ is represented by a tensor train operator. (b) The domain $[-W, W]$ is discretized using the quantics prescription, indexing exponentially fine increments with their binary representation. The number of tensors scales as $\log(\text{spatial resolution})$. (c) For a Gaussian wave packet $\psi(x) \propto \exp(-x^2/2)$ propagating in a modulated potential $\hat V(x) \propto  \sin^2(10x)+ \sin^2(10^5x)$ with a non-linear contribution $g=5$, we can resolve the wave function at all relevant length scales.}
    \label{fig:overview}
\end{figure}

Solving time-dependent non-linear differential equations computationally requires, in turn, some form of discretization of the spatial domain ($\boldsymbol{r}$) using finite elements, finite volumes or finite differences~\cite{LeVeque2007}, into $V$ points mapping the initial condition into a large vector of size $V$. This vector is then propagated in time using one of the many possible schemes that include Crank-Nicolson~\cite{Crank1947}, Fourier spectral~\cite{Feit1982}, Runge-Kutta~\cite{Runge1895} or Trotterization~\cite{Hatano2005} methods. These methods are typically limited in practice to $V\lesssim 10^9$, which can be a serious limitation in three dimensions and/or for problems that present vastly different characteristic length scales, which require the use of very fine grids. 
Recent methodological breakthroughs have shown that, for some problems, a numerical solution in a computational time scaling as fast as $\sim \log V$ was possible using a (compressed) tensor network representation~\cite{Learning_Feynman_TT, Fumega2024, Cross_extrapolation_low_rank, Tensorized_orbitals, QTCI_multivariate, learning_TT_noisy_functions, QTT_Feynman_multiorbital, compactness_QTT_propagators, Sun2025, 2025arXiv250605230A, non_eq_many_body_QTT, multiscale_space_time_QTT, quantics_HF, non_eq_GW_quantics, impurity_quantics, parquet_quantics, quantum_inspired_fluid, variational_excited, tensorizing_time, global_opt_qtt, TCI_purities, non_eq_Kondo, TCI_Bayesian_network, quantum_resource_TCI, entanglement_scaling_MPS, quantum_impurity_TCI, Chebyshev_quant_insp_numerics, Fokker_Planck, fast_flexible}. This representation has the dual quality of being very parsimonious while allowing one to perform many operations, including integration and differentiation, directly.
These methodological breakthroughs are made possible by the recent advent of learning algorithms such as the tensor cross-interpolation (TCI) algorithm~\cite{Oseledets2009, Oseledets2010, Oseledets2011,TCI_algorithms,Dolgov2012,Dolgov2020}. TCI allows one to efficiently transform the initial conditions and/or the various operators into the formalism of tensor networks.

In this paper, we demonstrate the application of the Quantics Tensor Train (QTT) representation for the numerical solution~\cite{PDE_quantics, num_sol_time_dep_GP, num_studies_split_step_fin_diff_non_linear_Schrödinger, numerical_sol_GP_BEC, time_dep_GP_anisotropic, eff_time_splitting_fin_diff_GP, num_sol_GP_trapped_BEC, GP_GPU, GP_Chia_Min} of the Gross-Pitaevskii (GP) equation~\cite{Gross1961, pitaevskii1961vortex}. Our primary goal is to study how the QTT approach performs in the presence of the additional difficulty associated with non-linearities. The GP equation describes the dynamics of Bose-Einstein condensates (BEC) or the modeling of confined particles in optical traps. In comparison to the Schrödinger equation, the GP equation exhibits an additional, "self-interaction" term, which introduces the third power of the wave function into the equation. Starting from a given initial state wave function, which is converted into a matrix product state (MPS), we perform a time evolution by performing a second-order Trotter evolution of the GP Hamiltonian. We take advantage of the fact that in the QTT representation, the Fourier transform can be performed exponentially faster than in the naive representation, allowing us to treat the kinetic energy directly in momentum space. The method is straightforward, easy to implement, and can be generalized to higher dimensions. We demonstrate it by solving the GP equation with spatial features ranging over seven orders of magnitude, requiring a domain discretization of about a billion increments. This can be achieved at roughly one millionth the memory cost of a comparable vectorized simulation and executed within hours on a conventional single-core desktop. Furthermore, we present two-dimensional solutions of the GP equation on a $2^{20}\times 2^{20}$ grid, thus with more than a trillion grid points.

This paper is organized as follows. We first introduce how QTTs and TCI can be applied to the numerical solution of time-dependent non-linear differential equations in Sec.~\ref{Sec_methods}. In Sec.~\ref{Sec_results}, we illustrate the algorithm with the simulation of BECs in modulated optical trap potentials in both one and two dimensions, including quasicrystalline potentials. Finally, we summarize our conclusions in Sec.~\ref{Sec_conclusion}.

\section{A semi-spectral integration approach}
\label{Sec_methods}

Here we show how the QTT representation can be naturally combined with a semi-spectral method to solve the non-linear problem.
For the sake of concreteness, we focus on the time-dependent Gross-Pitaevskii equation\cite{Gross1961,pitaevskii1961vortex}, noting that this approach can be directly extended to other non-linear equations. The time-dependent GP equation governs the mean-field dynamics of BECs\cite{RevModPhys.73.307}, and takes the (dimensionless) form
\begin{equation}
    i \frac{\partial \psi}{\partial t} = \left[ -\frac{1}{2m} \hat \nabla^2 + \hat V + g |\psi|^2 \right] \psi.
\end{equation}
Here, the wave function $\psi = \psi(\boldsymbol{r}, t)$ depends on space and time, and we assume the potential to depend on space only as $\hat V = \hat V(\boldsymbol{r})$. The parameter $g$ sets the self-interaction strength. The initial condition at $t=0$ is $\psi_0(\boldsymbol{r})$
defined in the domain $[-W,+W]^d$ in $d$ dimensions. Except for a few special cases, such as the soliton solution in one dimension~\cite{Dark_solitons_BEC}, the GP equation is not analytically solvable. 

Our strategy to integrate it is to use a Trotterization of its evolution using a small time step $h_t$, with the potential and interacting terms on one hand and the kinetic energy on the other hand. This choice is dictated by the fact that the corresponding evolution operators admit a low rank QTT representation respectively in real space (potential and interacting term) and momentum space (kinetic term) while one can use the superfast QTT Fourier transform \cite{QFT_small_entanglement} to navigate between the two. 

More precisely, we identify the different contributions to the dynamics as,
\begin{align}
    \hat H_\Delta &= -\frac{1}{2m} \boldsymbol{\hat \nabla}^2 \\
    \hat H_V &= \hat V(\boldsymbol{r}) \\
    \hat H_g(t) &= g |\psi(\boldsymbol{r},t)|^2
\end{align}
 and $\hat H = \hat H_\Delta + \hat H_V + \hat H_g$ is the total
 "Hamiltonian" (with a slight abuse of the terminology to include the non-linear term). Correspondingly, the evolution operator for a small time step $h_t$ is defined as
\begin{equation}
    \hat U_\xi (t,h_t) = e^{-i H_\xi(t) h_t} 
\end{equation}
with $\xi \in \{\Delta, V, g \}$ and simply $\hat U$ for the total Hamiltonian $\hat H$. The overall algorithm consists of the following steps:
\begin{enumerate}
    \item We define an exponentially fine spatial discretization step as $h_r=2W/2^R$ where $Rd$ will be the number of indices of our MPS.
    We further define the time discretization step  $h_t << T$, where $T$ is the total evolution time.
    \item We use TCI to transform $\psi_0(\boldsymbol{r})$, $\hat U_V$ and $\hat U_K\equiv e^{-ik^2 h_t/(2m)}$ into a QTT form. We further construct the matrix product operator representations of the fast Fourier transform~\cite{multiscale_space_time_QTT} $\hat U_{\text{FFT}}$.
    \item For each of the $T/h_t$ time steps, we obtain
    \begin{equation}
        \ket{\psi(t+h_t)} = \hat U(t) \,\ket{\psi(t)}
    \end{equation}
    by applying alternatively $\hat U_V \hat U_g$ and $\hat U_\Delta =\hat U_{\text{FFT}}^{-1} \hat U_K \hat U_{\text{FFT}}$ on the state. 
    \item In each time step, the non-linear evolution $\hat U_g\ket{\psi(t)}$ is constructed using TCI for the object $\exp(-ig |\psi(\boldsymbol{r},t)|^2 h_t) \psi(\boldsymbol{r},t)$.
\end{enumerate}

This strategy is illustrated in Fig.~\ref{fig:overview}, where~\ref{fig:overview}a visualizes the time evolution expressed through a repeated application of matrix product operators (MPO). Fig.~\ref{fig:overview}b shows how an initial condition $\psi_0(\boldsymbol{r})$ and a given potential $\hat V(\boldsymbol{r})$ can be indexed in real space via the quantics representation, which associates given increments in space with a $R$-adic binary index. Each additional subdivision in length scale results in an additional tensor added to the corresponding MPO and QTT representations, shown by the corresponding color codings in Figs.~\ref{fig:overview}a,b. In Fig.~\ref{fig:overview}c, we show an example of a time evolution of a Gaussian wave packet $\psi_0(x) = 1/\sqrt{N} \exp(-x^2/2)$ in a doubly-modulated potential $\hat V(x) \propto  \sin^2(10x)+ \sin^2(10^5x)$, using $R=30$. Two zooms highlight periodic modulations in the probability density at the different relevant length scales, showing that we can resolve the wave function in the entire domain.

\begin{figure}[t!]
    \centering
    \includegraphics[width=\columnwidth]{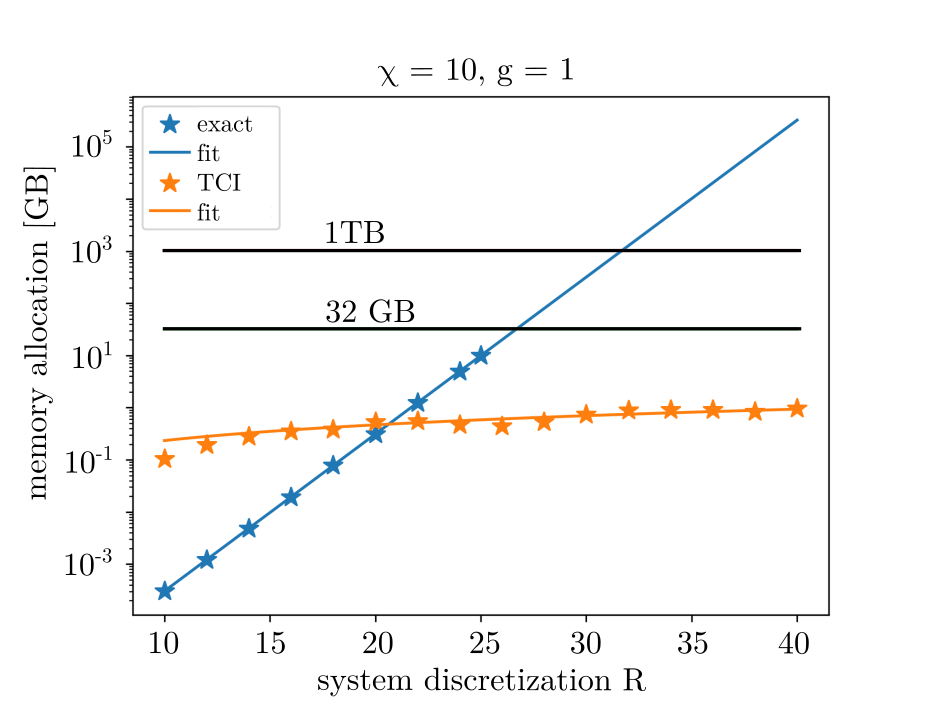}
    \caption{Memory allocation per Trotter step, compared for a vectorized simulation (ED) and the TCI simulation. For system discretizations greater than $R \approx 22$, TCI outperforms any exact vectorized approach.}
    \label{fig:memory_allocation}
\end{figure}

In Fig.~\ref{fig:memory_allocation}, we compare the required memory allocation per Trotter step for our algorithm based on QTCI (for a given bond dimension $\chi=10$) with an exact vectorized (ED) finite-difference implementation of the same algorithm, using a conventional fast Fourier transform and element-wise multiplications in Julia. With an increase in the number of increments, the required memory grows exponentially in $R$ for ED, whereas it grows only linearly with $R$ for TCI. For system sizes greater than roughly $R=22$, the TCI algorithm outperforms the vectorised implementation.

\subsection{Trotterized solution of Gross-Pitaevskii equation}

We now elaborate on the details and technical aspects of the Trotterized time evolution of the GP equation. 
In general, to solve the GP equation numerically, we seek a finite-difference approximation of the time evolution operator $\hat U(t)$ in a box $[-W, W]^d$, with $L=2^R$ spatial increments such that $h_r = 2W/L$. We implement this time-evolution operator by Trotterization up to second order. Although other numerical methods exist to solve such partial differential equations, such as the Crank-Nicolson or Runge-Kutta algorithms, Trotterization has the advantage of being manifestly unitary and unconditionally stable at every step of the computation. Furthermore, it is not necessary to perform operator inversions as in the Crank-Nicolson method. 

The second-order Trotter approximation of $\exp(-i \hat H h_t)$ for a small time increment $h_t$ splits $\hat H$ into two non-commuting parts $\hat H_\Delta$ on one hand and $\hat H_V$ and $\hat H_g$ on the other hand.
It reads,
\begin{equation}
    \hat U(h_t) = \hat U_{V+g} \left(\frac{h_t}{2} \right)  \hat U_{\Delta} \left(h_t \right) \, \hat U_{V+g} \left(\frac{h_t}{2} \right)  + \mathcal{O}(h_t ^3).
\end{equation}
 Choosing time increments on the order of $h_t = 10^{-2} - 10^{-3}$ will typically yield acceptably low overall errors. To construct a diagonal finite difference representation of the kinetic time-evolution operator $\hat U_{\Delta}$, we transform the operator to momentum space as $\hat U_{\Delta} (t) =  \hat U_{\text{FFT}}^{-1}  \exp(-i \hat H_{K} t) \, \hat U_{\text{FFT}} $, with 
\begin{equation}
\label{fin_diff}
    \hat H_K = \frac{2 L^2}{m(2W)^2} \sum_{\boldsymbol{k}=0}^{L-1} \sum_{i=1}^d \sin^2 \left(  \frac{\pi k_i}{L} \right) \ket{\boldsymbol{k}} \bra{\boldsymbol{k}}
\end{equation}
and where $\hat U_{\text{FFT}}$ and $\hat U_{\text{FFT}}^{-1}$ denote the fast Fourier transform operator and its inverse. The complete Trotterized time-evolution operator is then given by the product
\begin{equation}
    \begin{split}
    \ket{\psi(T) } =& \\
        &\hat U_V \left(\frac{h_t}{2} \right) \hat U_g \left(\frac{h_t}{2} \right) \hat U_{\text{FFT}}^{-1} \hat U_K \left(h_t\right) \hat U_{\text{FFT}} \\
        &\left( \prod_{i=1}^{N-1} \hat U_V \left(h_t \right) \hat U_g \left(h_t \right) \hat U_{\text{FFT}}^{-1} \hat U_K \left(h_t\right) \hat U_{\text{FFT}}\right) \\
        &\hat U_V \left(\frac{h_t}{2} \right) \hat U_g \left(\frac{h_t}{2} \right)\ket{\psi_0}
    \end{split}
\end{equation}

\subsection{Quantics tensor cross interpolation}

\begin{figure*}[t!]
    \centering
    \includegraphics[width=0.99\textwidth]{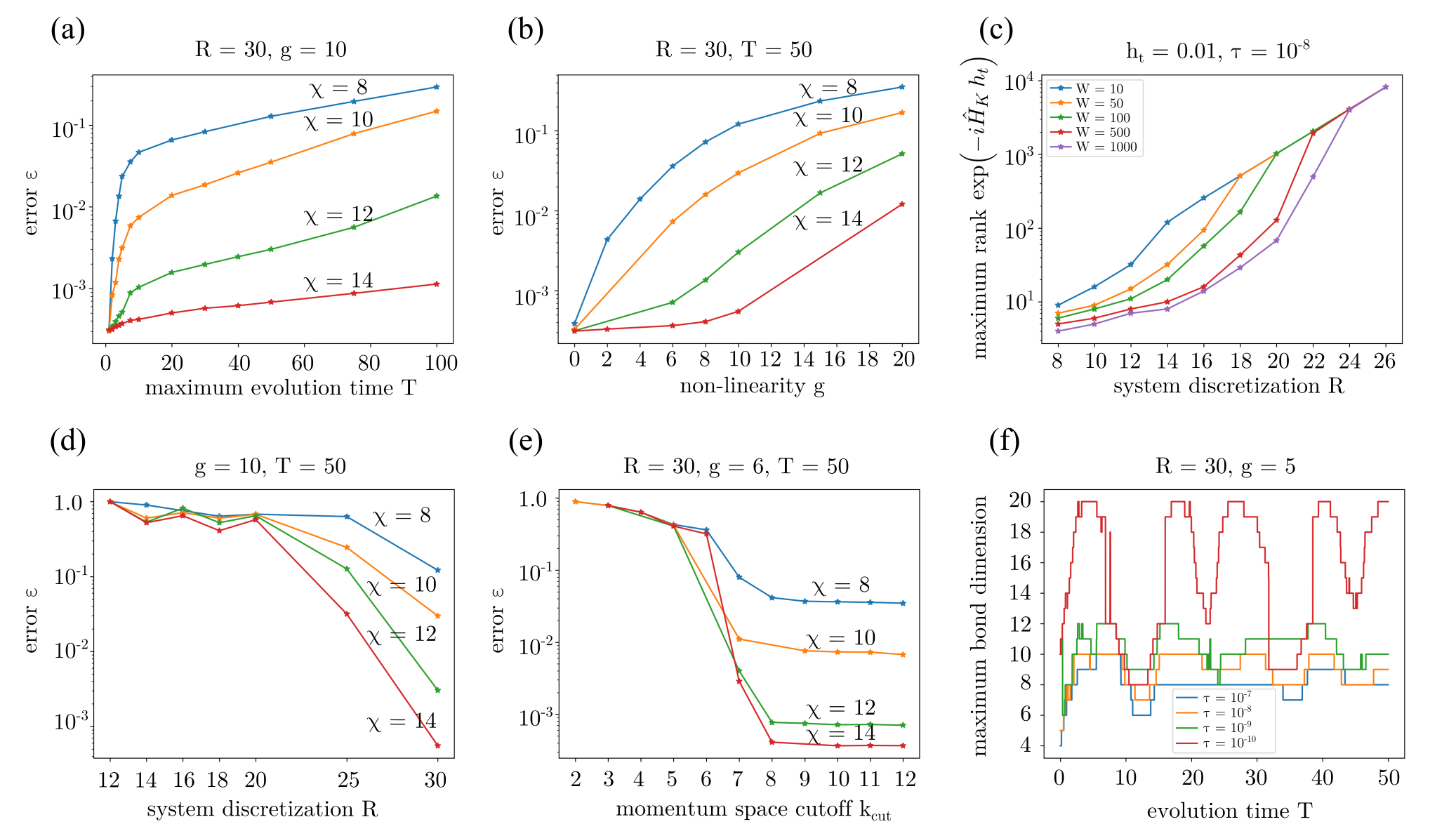}
    \caption{Performance metrics of the mixed-spectral Trotterized solution of the Gross-Pitaevskii equation, using quantics tensor trains. We time-evolve a Gaussian wave packet $\psi_0(x) \propto \exp(-x^2/2)$ in a modulated optical trap $\hat V(x) = 0.01x^2 + 10\sin^2 (10^4 x)$ with time increment $h_t=0.01$. (a, b) Infidelity $\epsilon$ of a forward-backward time-evolved state $\ket{\psi_{\text{evol}}}$ with respect to the initial state $\ket{\psi_0}$. For larger evolution times $T$ and non-linearities $g$, the error monotonously increases. Higher cutoffs in the bond dimension reduce the error. (c) Maximum rank of the momentum-space Laplacian $\exp(-i \hat H_K h_t)$, as a function of the discretization $R$. For $R \gtrsim 20$, the rank quickly grows to $\mathcal{O}(1000)$. (d) To describe all relevant features of this given benchmark system, a number of increments $2^R$ much greater than the extent of all involved length scales is needed. (e) To control the rank of the momentum space Laplacian, a low-pass filter with a cutoff keeping $2^8$ momenta is introduced.(f) Maximum bond dimension per time step for different given tolerances. There is no net increase of the bond dimension with time.}
    \label{fig:forward_backward_1}
\end{figure*}

To accurately resolve phenomena ranging across multiple length scales, we require a grid resolution $h_r$ much smaller than the smallest involved length scale $\lambda$. When $W/\lambda \gg 1$ is large, this becomes a problem for conventional discretization schemes. An illustration of this problem was shown in Fig.~\ref{fig:overview} where $W=200$ but $\lambda\approx 10^{-5}$
(the oscillations of the potential) so that $W/h_r\approx 10^9$ discretization points needed to be used. The QTT representation, described now, solves this problem.

First,  we summarize the idea behind QTT in one dimension; the higher-dimensional version works analogously. By choosing an exponentially fine discretization of the domain $[-W, W]$ as $h_r = 2W/2^R$, we may identify each point either by a linear index, $x(m) = -W + m h_r$ with $m = 0, 1, ..., 2^R-1$, or by the binary representation of $m = \sum_{i=1}^R \sigma_i 2^i \equiv (\sigma_R \sigma_{R-1} ... \sigma_1)$. Correspondingly, each entry $\psi_m = \psi(-W + m h_r)$ of a discretized wave function $\psi(x)$ can be identified by the binary expansion of $m$, as $\psi_m = \psi(-W + \sum_{i=1}^R \sigma_i 2^i \, h_r) \equiv \psi(\sigma_R \sigma_{R-1} ... \sigma_1)$. This final step defines an effective rank-$R$ tensor that has $R$ legs of local dimension two. Typically, quantities such as the wave functions that are solutions to equations of physics are highly structured objects, and therefore, the tensor $\psi(\sigma_1 \sigma_2 \, ... \sigma_R)$ often exhibits a high degree of compressibility~\cite{QTCI_multivariate}. Using the TCI algorithm~\cite{TCI_algorithms}, we obtain a quasi-optimal
QTT approximation of the wave function as 
\begin{equation}
    \psi(\sigma_R \sigma_{R-1} ... \sigma_1) \approx M_{\sigma_R}^{i_{R-1}} M_{\sigma_{R-1}}^{i_{R-1} i_{R-2}} ... \, M_{\sigma_1}^{i_1}. 
\end{equation}
Here, the $a^\text{th}$ tensor contains the information describing the function $\psi(x)$ at length scale $2^{-a}$. The "virtual" indices $\{i_1, i_2, ... i_{R-1} \}$ are understood to be summed over (Einstein implicit summation convention).  Each virtual index $i_a$ has dimension $\chi_a$
(the "bond dimension") which controls the degree of entanglement between different length scales.

The MPO representation of the different building blocks of the time-evolution operator $\hat U$ can be obtained analogously. For a given potential $\hat V(x)$, we provide the function $U_V (x) =  \exp(-i \hat V(x) h_t)$ to the TCI algorithm, yielding a QTT $(U_V)_{\tau_R \tau_{R-1} ... \tau_1}$. This QTT $U_V$ can then be promoted to a diagonal operator by performing an element-wise multiplication with the QTT $\psi_{\sigma_R \sigma_{R-1} ... \sigma_1}$ representing the wave function. Formally, this can be achieved by introducing a summation over an auxiliary index defined by 3-dimensional delta-functions:
\begin{equation}
    \psi'_{\rho_R ... \rho_1} = (U_V)_{\tau_R  ... \tau_1} \,
    \delta_{\tau_R \sigma_R \rho_R} \,
    ... \,
    \delta_{\tau_1 \sigma_1 \rho_1} \,
    \psi_{\sigma_R  ... \sigma_1}.
\end{equation}

In cases where the potential $\hat V(x)$ is given by a sum of terms $\hat V(x) = \sum_i \hat V_i(x)$, such as in small-scale modulations on top of a large-scale trapping potential, it is advantageous to construct the MPO representations $U_{V_i}$ of each potential evolution term $\hat U_{V_i}$ separately. This incurs no additional Trotter errors, as $[\hat V_i(x), \hat V_j(x)] = 0$. The overall potential evolution is then given by the product $\hat U_{V} \, \ket{\psi(t)}=\prod_i \hat U_{V_i} \, \ket{\psi(t)}$.

Finally, to implement the non-linear transformation
\begin{equation}
    \psi(x) \rightarrow \exp(-ig |\psi(x)|^2 h_t) \psi(x) \equiv f(\psi(x)) \equiv \tilde f(x),
\end{equation}
we apply TCI to the function $\tilde f(x)$ defined above. Here, for each pivot $x$ sampled during the TCI algorithm, the corresponding value of $\psi(x)$ at a given time step is evaluated to obtain the value of the function $\tilde f(x)$. It is therefore not necessary to construct a separate QTT of the operator $\hat U_g$. 

\begin{figure*}[t!]
    \centering
    \includegraphics[width=0.98\textwidth]{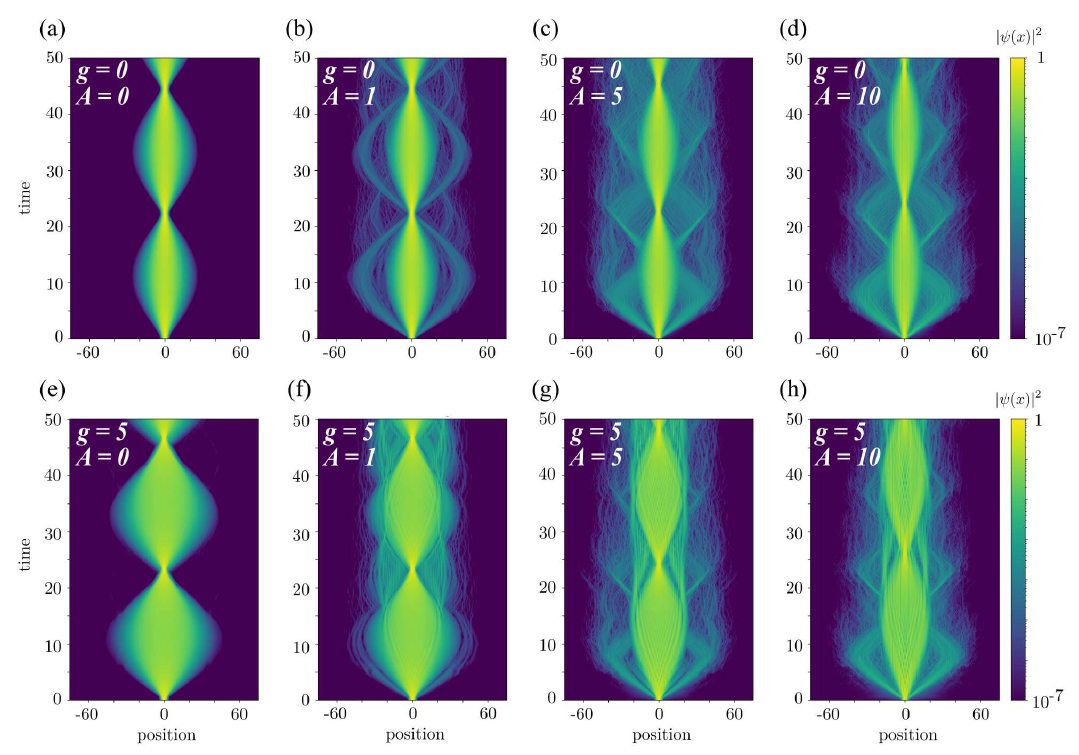}
    \caption{Time evolution of a Gaussian wave packet $\psi(x) = (1/\pi)^{1/4} \exp (-x^2/2)$ in the potential $\hat V(x) = A_x/W^2 x^2 + A \sin^2(q x)$, with $A_x/W^2 = 0.01$ and $q = 3$. We show different values of the nonlinearity $g$ and the modulation amplitude $A$. We perform a time evolution up to $T=50$ with $h_t=0.01$ and bond dimension $\chi=10$.}
    \label{fig:quad_mod_evolution_multipanel_small}
\end{figure*}

\subsection{Kinetic evolution and low-pass filter in momentum space}

We now discuss the details of the implementation of the kinetic energy operator with tensor networks and QTCI. We implement the kinetic time evolution operator in momentum space, transforming $\psi(x, t)$ into momentum space via a Fourier transform. The Fourier transform operator $\hat U_{\text{FFT}}$ can be exactly implemented as a discrete Fourier transform in QTT format as 
\begin{equation}
    (U_{\text{FFT}})_{\boldsymbol{\sigma'} \boldsymbol{\sigma}} = \frac{1}{\sqrt{2^R}} \exp \left( -i 2\pi \sum_{l, l'=1} ^{2^R} 2^{R-l-l'} \sigma_{l'} \sigma_l \right),
\end{equation}
where $\boldsymbol{\sigma'}$ and $ \boldsymbol{\sigma}$ are the corresponding quantics indices, where the MPO $U_{\text{FFT}}$ is of low rank~\cite{QFT_small_entanglement}. It is important to note that the momentum space indices are in reverse order with respect to the position space indices, which can be intuitively understood as an instance of Fourier reciprocity - small distances correspond to high frequencies and vice versa. This ordering needs to be accounted for when constructing operators in momentum space. 
This allows us to implement the kinetic evolution operator $\hat U_{\Delta}$ as a diagonal operator in momentum space via $\hat U_{\Delta} = \hat U_{\text{FFT}}^{-1} \hat U_{K} \hat U_{\text{FFT}}$, where $\hat U_{K}$ is the exponential of the finite-difference Laplacian operator of Eq.~\ref{fin_diff}. However, the maximum rank of $\hat U_{K}$ increases rapidly with increasing resolution $R$, as shown in Fig.~\ref{fig:forward_backward_1}c. To reach resolutions of $R>20$ in practice, it is necessary to compress the operator to a lower rank before applying it to a state. This can be achieved by defining a low-pass operator, which effectively cuts out the fast-oscillating high-frequency components of $\hat U_{K}$. We implement this cutoff in the kinetic energy through a softened step function of the form
\begin{equation}
    \hat \Theta(\boldsymbol{k}; k_{\text{cut}}, \beta) = \prod_{i=1}^d \frac{1}{\exp[(k_i-k_{\text{cut}})\beta] + 1},
\end{equation}
Here, the parameter $\beta$ regulates the sharpness of the jump and can be adjusted to smooth out the cutoff value. In our computations, we have found $\beta = 2$ to be a suitable choice. In practice, by computing the QTT representation $(\Theta \, U_K)$ of the product $\hat \Theta(\boldsymbol{k}; k_{\text{cut}}, \beta) \hat U_{K}$
the maximum rank of $U_K$ can be reduced from $\mathcal{O}(100-1000)$ to less than $\mathcal{O}(10)$. In each computation, it is then most efficient to construct the whole kinetic MPO
\begin{equation}
    U_{\Delta} \equiv U_{\text{FFT}}^{-1} \, (\Theta \, U_{K}) \, U_{\text{FFT}}
\end{equation}
as a single operator, which typically has a maximum bond dimension of less than $20$. This MPO only needs to be computed once, and can then be re-used in every time step of the Trotter evolution. Correspondingly, $U_{\Delta}$ can be constructed with high accuracy by imposing low relative tolerance cutoffs on the order of $10^{-10}$, as this only leads to an inconsequential computational overhead in the entire time-evolution. We have observed a better accuracy in the operator construction when using a naive MPO-MPO contraction, compared to other contraction algorithms such as the zipup-contraction~\cite{Paeckel2019}. For the construction of $(\Theta \, U_{K})$, we have found that choosing a momentum space cutoff $k_{\text{cut}}$ such that the $2^8$ lowest momenta are kept is typically sufficient for high precision (see Fig.~\ref{fig:forward_backward_1}e). To protect against convergence issues, we set the momentum space cutoff dynamically during the MPO construction of $(\Theta \, U_{K})$ by computing a single kinetic evolution step of a trial wave packet. In practice, if $k_{\text{cut}}$ is chosen too low, it will lead to a drastic non-conservation of wave function norm. Therefore, we monitor the norm of the trial wave packet and increase $k_{\text{cut}}$ iteratively until we observe approximate norm conservation, which typically happens for $k_{\text{cut}} = 2^8$ or $2^9$.

\section{Illustrative numerical simulations}
\label{Sec_results}

\begin{figure*}[t!]
    \centering
    \includegraphics[width=0.98\textwidth]{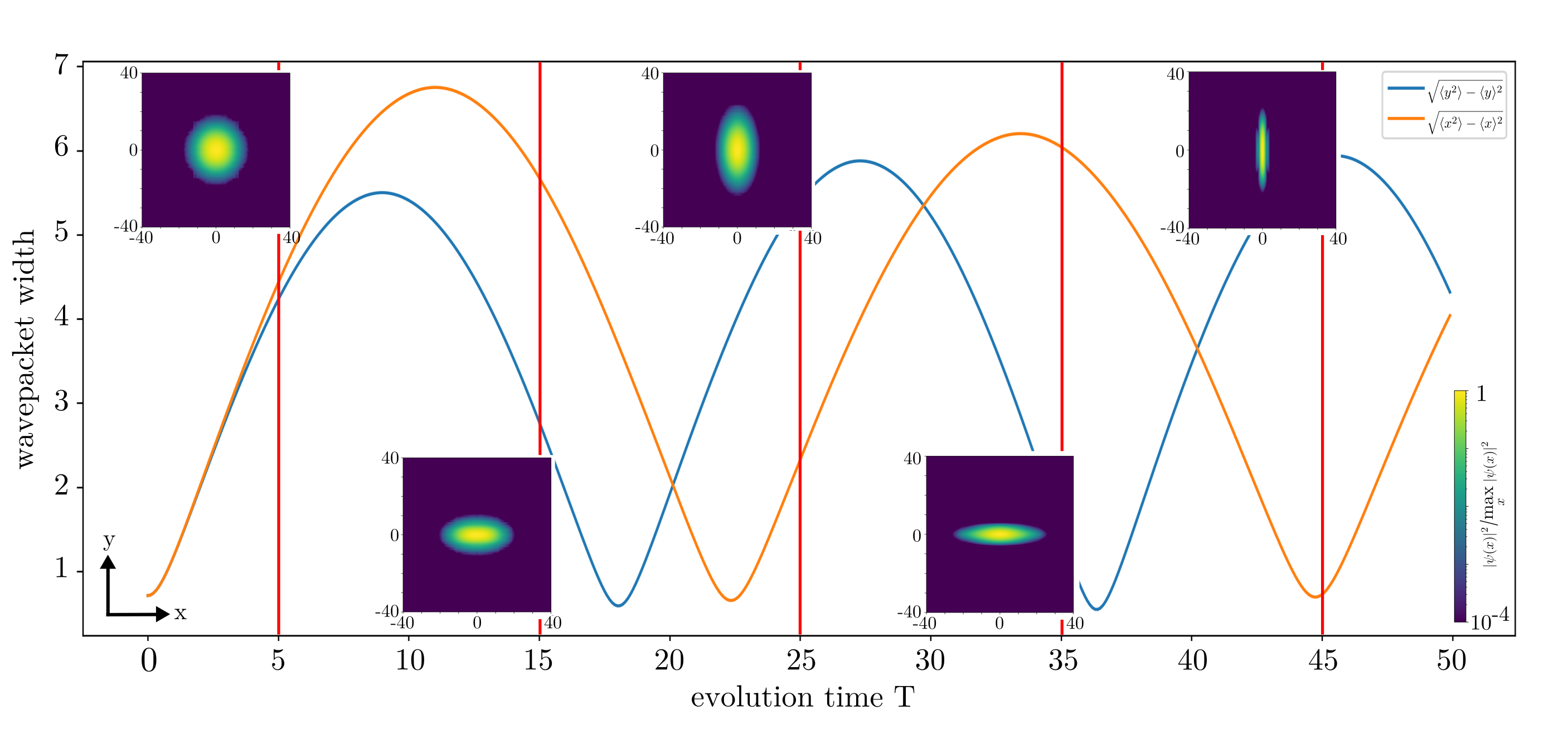}
    \caption{Time evolution of a two-dimensional BEC, represented by a Gaussian wave packet $\psi_0(x, y) \propto \exp(-(x^2 + y^2)/2)$, in an anisotropic harmonic potential $\hat V(x, y)= 0.01 x^2 + 0.015 y^2$. We show the width $\sqrt{\langle x^2 \rangle - \langle x \rangle^2}$ ($\sqrt{\langle y^2 \rangle - \langle y \rangle^2}$) of the wave packet in $x$ ($y$) direction as a function of evolution time $T$. The insets show the probability density in space at selected times. We use a grid resolution of $2^{20}\times 2^{20}$ increments in a domain $[-100, 100]^2$, modelled by a tensor train of length $40$, and time increments $h_t=0.01$. The strength of the non-linearity is $g=5$. For spatial evolutions, we impose a maximum bond dimension of $50$ and for the non-linear evolution a tolerance of $10^{-8}$.}
    \label{fig:2D_test}
\end{figure*}

For all the calculations discussed here~\cite{GP_TCI}, we have employed the ITensors\cite{itensor}, QuanticsTCI\cite{QuanticsTCI.jl}
and TensorCrossInterpolation\cite{TensorCrossInterpolation.jl} packages in the Julia language. The overall discretization $R$ of the computation should be set such that all involved length scales can be resolved, meaning that $2W/2^R = h_r \ll \lambda$, where $\lambda$ describes the spatial separation of features at the smallest involved length scale. Taking $h_r = 10^{-2} \lambda$ typically yields satisfactory results. The MPO $U_\Delta$ representing the kinetic evolution term with a low-pass filter in momentum space does not depend strongly on $R$ and has a bond dimension of around 20. The MPO $U_V$ representing the potential evolution term $\hat U_V$ is, however, dependent on the exact functional shape of $\hat V(x)$ and may reach bond dimensions up to $\mathcal{O}(100)$. We have observed that, to obtain a well-converged tensor train representation $U_V$ of $\hat U_V$, the time increment $h_t$ should be chosen such that $|\hat V(x) h_t| \lesssim 1$. To implement the MPO-MPS contraction for the potential evolution, we utilize a variational algorithm that finds the optimal compressed MPS in a manifold of a specified bond dimension by executing DMRG-like sweeps~\cite{Paeckel2019},
which has a two-fold advantage. First, it has a lower cost in the bond dimension of the MPO, which makes it suitable for larger potential MPOs. Second, it automatically takes the previous MPS as an input state for the variational algorithm, which is especially convenient as the final state has a high overlap with the initial state. Typically, a bond dimension comparable to the initial state can be imposed throughout the entire computation. Throughout the computation, the norm is not exactly conserved as a consequence of the momentum-space low-pass filter in the kinetic evolution. We therefore renormalise the state at every step of the time evolution.

\subsection{Modulated optical trap potentials in one dimension}

As a first example of our method, we simulate a modulated harmonic potential, defined as 

\begin{equation}
    \hat V(x) = A_x \left(\frac{x}{W}\right)^2 + A\sin^2(q\, x).
\end{equation}
By defining different modulation frequencies with the wave vector $q$, we can change the smallest length scale at which the potential has non-trivial features. The dynamics in this potential correspond to dynamics of BECs in optical traps, which can be realized experimentally in cold atom setups\cite{Anglin2002,Greiner2002,Schreiber2015,RevModPhys.85.553}. As the initial state $\ket{\psi_0}$, we take the ground state of a quantum harmonic oscillator with $A_x/W^2=1$, represented by the wave function
\begin{equation}
    \psi_0(x) = \left( \frac{1}{\pi} \right)^{1/4} e^{-x^2/2}.
\end{equation}
To benchmark our algorithm, we compute the entire evolution up to time $T$, then compute it backward in time all the way back to $t=0$ (using $h_t\rightarrow -h_t$). After this entire forward-backward evolution, we obtain the state $\ket{\psi_{\text{evol}}}$ which, supposingly, should be equal to $\ket{\psi_0}$. We compute the error $\epsilon$ of the time-evolved state with respect to the initial state as
\begin{equation}
    \epsilon = 1 - \vert \langle \psi_{\text{evol}} \ket{\psi_0}  \vert^2.
\end{equation}
The results for an optical trap potential with a modulation of $q = 10^4$, requiring precision over a range of six orders of magnitude, are shown in Fig.~\ref{fig:forward_backward_1}. 

In Figs.~\ref{fig:forward_backward_1}a,b,d,e, we show the error $\epsilon$ incurred in a forward-backward time evolution as a function of different system parameters. As expected, the error $\epsilon$ grows with increasing evolution time, reflecting a build-up of both the Trotter and TCI errors, and is lower for greater bond dimensions. Even for evolution times up to $T=100$, the error remains at the per mill level. Furthermore, $\epsilon$ grows with increasing the non-linear contribution $g$. To resolve the features of the test system, ranging across six orders of magnitude satisfactorily, we need to employ system discretization considerably larger than $10^6 \approx 2^{20}$, as shown in Fig.~\ref{fig:forward_backward_1}d. Furthermore, it is generally sufficient to keep the lowest $2^8$ momentum in the construction of the kinetic evolution MPO in momentum space, as for higher numbers of momenta kept, the error remains in practice unaffected.

In all the simulations discussed above, the error $\epsilon$ can be lowered by increasing the maximum allowed bond dimension for the QTT representation of $\psi(x,t)$ at a given time slice. We observe that in this case, a bond dimension of $\chi=10$ yields a satisfactory precision. The memory needed for the QTT describing the BEC is therefore upper bounded by a fraction of $30\cdot 10^2 \cdot 2/2^{30} \approx 5.6 \cdot 10^{-4} \%$, compared to a conventional vectorised finite-difference simulation. Finally, in Fig.~\ref{fig:forward_backward_1}f, we show the maximum bond dimension at each time step for different given per-tensor error tolerances $\tau$ in the TCI and MPO-MPS contraction algorithms. Here, $\tau$ is to be understood as the normalized approximation error of a given tensor $M$ in the QTT representation of $\psi(x)$, as $\tau = || \tilde M - M||/||M||$ with $\tilde M$ the approximated tensor. Apart from period fluctuations, after an initial build-up, the bond dimension does not increase with time. This is due to the fact that the BEC wave function undergoes a periodic evolution where the features at different length scales, once built up by tunnelling through the wells defined by the fast-oscillating modulation, do not drastically change with time. This should be contrasted with conventional MPS time evolution algorithms such as time-dependent variational principle (TDVP) or time-evolving block-decimation (TEBD) in the context of quantum many-body problems, where the MPS rank grows with time, making the time evolution either computationally more expensive or more inaccurate in the case of truncations for large evolution times $T$\cite{Paeckel2019,PhysRevB.94.165116}.

\begin{figure*}[t!]
    \centering
    \includegraphics[width=0.98\textwidth]{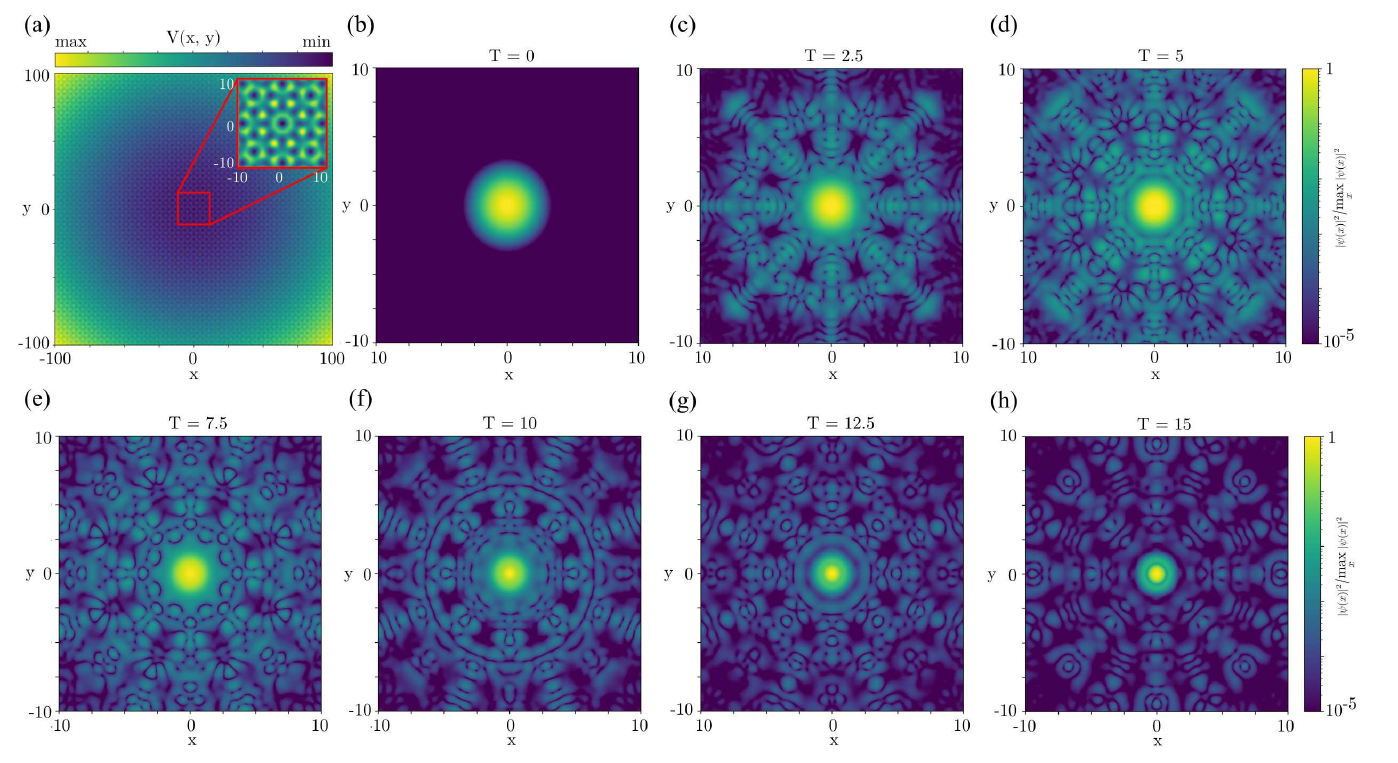}
    \caption{Time evolution of a two-dimensional BEC, represented by a Gaussian wave packet $\psi_0(x, y) \propto \exp(-(x^2 + y^2)/2)$, in a harmonic potential with eightfold symmetric sinusoidal modulations $\hat V(x, y)= 0.01 (x^2 + y^2) + 10 \sum_{i=1}^4 \sin^2(\boldsymbol{q_i}\cdot \boldsymbol{r})$. We use a grid resolution of $2^{20}\times 2^{20}$ increments in a domain $[-100, 100]\times [-100, 100]$, modelled by a tensor train of length $40$, and time increments $h_t=0.01$. The strength of the non-linearity is $g=5$. For spatial evolutions, we impose a maximum bond dimension of $50$ and for the non-linear evolution a tolerance of $10^{-8}$.}
    \label{fig:2D_mod}
\end{figure*}

We show several examples of the BEC evolution in the modulated optical trap potential for different parameter values in Fig.~\ref{fig:quad_mod_evolution_multipanel_small}. The top row displays the case of the non-linearity $g$ set to $0$, i.e. a Schrödinger evolution, and the bottom row displays a GP evolution with $g=5$. From left to right, we increase the amplitude of the modulation from $A=0$ to $A=10$. The optical trap is modulated at a frequency of $q = 3$, creating potential troughs and crests of a width roughly equal to $1$. In all cases, switching on the non-linearity leads to an effective broadening of the BEC's breathing, while the breathing frequency in time remains unaffected. A non-trivial modulation of the harmonic trap potential leads to two effects. First, the width of the BEC breathing becomes effectively smaller, as there are additional potential crests that prevent the wave function from exploring a range as wide as in the non-modulated case. Second, the wave function will tunnel through the potential crests closest to the initial state $\psi_0(x)$, which leads to a smearing of the probability amplitude around the BEC.

\subsection{BEC evolution in two-dimensional optical traps}

\begin{figure*}[t!]
    \centering
    \includegraphics[width=\textwidth]{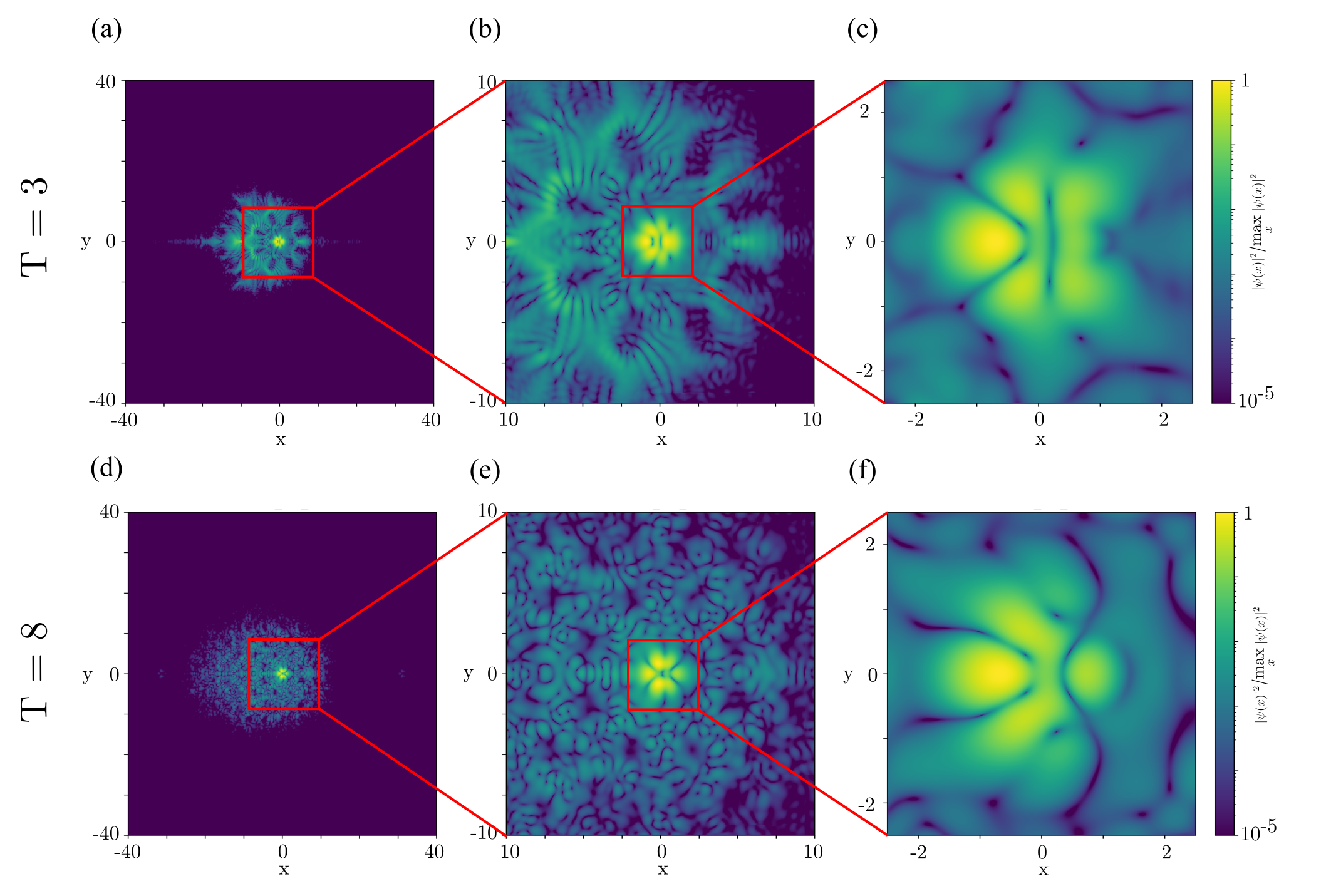}
    \caption{Time evolution of a two-dimensional BEC with initial velocity inn x-direction, represented by a Gaussian wave packet $\psi_0(x, y) \propto \exp(-(x^2 + y^2)/2)\cdot e^{ikx}$ with $k=5$, in a harmonic potential with eightfold symmetric sinusoidal modulations $\hat V(x, y)= 0.01 (x^2 + y^2) + 10 \sum_{i=1}^4 \sin^2(\boldsymbol{q_i}\cdot \boldsymbol{r})$. We use a grid resolution of $2^{20}\times 2^{20}$ increments in a domain $[-100, 100]\times [-100, 100]$, modelled by a tensor train of length $40$, and time increments $h_t=0.01$. The strength of the non-linearity is $g=5$. For spatial evolutions, we impose a maximum bond dimension of $50$ and for the non-linear evolution a tolerance of $10^{-8}$. The two rows show the probability density at selected times $T=3$ and $T=8$ with a progressive zoom into smaller-scale features.}
    \label{fig:2D_eigthfold_moving}
\end{figure*}

In the following, we illustrate the application of QTCI methods to solve non-linear partial differential equations in two dimensions, which is a crucial step forward in comparison to conventional, vectorized simulations. For concreteness, we will again consider the evolution of a BEC represented by a Gaussian wave packet
\begin{equation}
    \psi_0(x, y) = \frac{1}{\sqrt{N}} e^{-(x^2 + y^2)/2}
\end{equation}
in different modulated optical trap potentials
\begin{equation}
\begin{split}
    \hat V(x, y) = \, &A_x \left(\frac{x}{W}\right)^2 + A_y \left(\frac{y}{W}\right)^2 + A_{xy} \frac{xy}{W^2} \\
    + &A\sum_{i} \sin^2(\boldsymbol{q_i} \cdot \boldsymbol{r}),
\end{split}
\end{equation}
with $\boldsymbol{r} = (x, y)$. By adjusting the prefactors $A_x/W^2, A_y/W^2$ and $A_{xy}/W^2$, we can model isotropic and anisotropic potential traps, which may be modulated by a finer grid of sine functions of amplitude $A$, defined by their respective wave numbers $\boldsymbol{q_i}$. With conventional methods, the domain $[-W, W]^2$ can be discretized into roughly $2^{10} \times 2^{10} \approx 1 \,\text{million}$ increments before the computational cost becomes prohibitively large, which severely restricts the attainable spatial resolution in both $x$- and $y$-directions. In the following calculations, we restrict the maximum bond dimension of the potential evolution operator $U_V$ to $50$ in each evolution step, but allow the bond dimension of the non-linear evolution to grow until a tolerance of $10^{-8}$ is reached. 

We first consider a minimal two-dimensional case, namely a BEC in an anisotropic potential. In Fig.~\ref{fig:2D_test}, we show the evolution of a Gaussian wave packet as defined above in an anisotropic harmonic potential $\hat V(x, y) = 0.01 x^2 + 0.015 y^2$. The anisotropy is reflected in the wave packet widths in $x$- and $y$-direction, which differ in both amplitude and frequency. The BEC therefore breathes with different periodicities in the $x$- and $ y$-directions, as shown in the insets at selected time steps.

We now consider the dynamics of a BEC in an eightfold symmetric quasicrystalline potential, a setup recently demonstrated experimentally\cite{PhysRevLett.122.110404, Yu2024}.
In Fig.~\ref{fig:2D_mod}, we show the evolution of a BEC in a modulated optical trap with eightfold rotational symmetry, defined by the four wave vectors
\begin{equation*}
    \boldsymbol{q_1} = (1, 0), \, \, \boldsymbol{q_2} = \frac{1}{\sqrt{2}}(1, 1),  \, \, \boldsymbol{q_3} = \frac{1}{\sqrt{2}}(-1, 1),  \, \, \boldsymbol{q_4} = (0, 1).
\end{equation*}
We take an isotropic harmonic trapping potential with $A_x/W^2 = A_y/W^2 = 0.01$ and a modulation amplitude of $A=10$. The potential is visualized at two different length scales in Fig.~\ref{fig:2D_mod}a, showing both the large-scale trap in the entire domain as well as the short-range modulations in a restricted domain $[-10, 10] \times [-10, 10]$ in the inset. Figs.~\ref{fig:2D_mod}b-h then show the probability density of the two-dimensional wave function at selected times during the time evolution, up to a maximum time of $T=15$. We have again used time increments $h_t = 0.01$ and the strength of the non-linearity $g$ is set to $5$. At all times shown, the BEC tends to remain localized in the central potential well, surrounded by eight peaks of the quasicrystalline modulation. With increasing time, the probability density leaks into adjacent wells, following the eight-fold symmetry defined by the sinusoidal potential landscape. The maximum spread is visible around the time steps shown at $T=5$ and $T=7.5$. After that, the BEC starts retracting towards the center of the domain, with the leakage pattern becoming increasingly faint.

In Fig.~\ref{fig:2D_eigthfold_moving}, we show the time evolution of a BEC with non-zero initial velocity in the same eightfold symmetric quasicrystalline potential. 
The initial state is now defined as a Gaussian wave packet times a plane wave,
\begin{equation*}
    \psi_0(x, y) = \frac{1}{\sqrt{N}} e^{-(x^2 + y^2)/2} \, e^{ikx}.
\end{equation*}
In contrast with the case considered in  Fig.~\ref{fig:2D_mod}, we now take a finite wave number $k=5$, which leads to finite momentum in the BEC.
The top and bottom rows of Fig.~\ref{fig:2D_eigthfold_moving} show two selected points in the time evolution, displaying the probability density at $T=3$ in Figs.~\ref{fig:2D_eigthfold_moving}a-c and at $T=8$ in Figs.~\ref{fig:2D_eigthfold_moving}d-f, respectively. From left to right, we zoom progressively into the domain. As in Fig.~\ref{fig:2D_mod}, we observe a modulation of the probability density following the modulation of the trapping potential, but with an additional displacement due to the finite initial state velocity, leading to the BEC exploring regions further to the left. 

\begin{figure}[t!]
    \centering
    \includegraphics[width=\columnwidth]{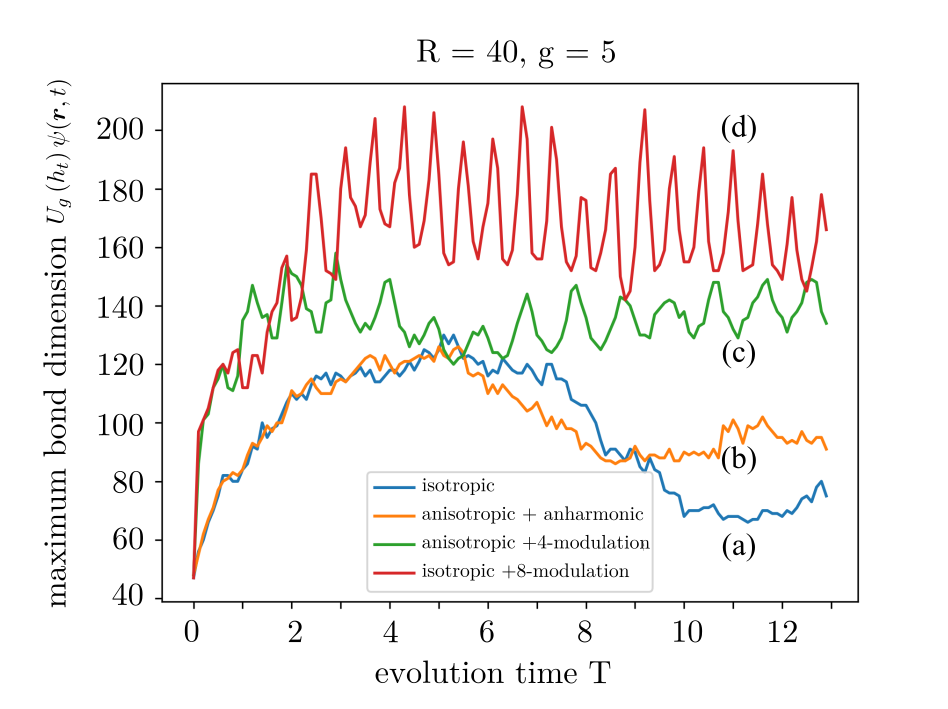}
    \caption{Maximum bond dimension of $U_g \, \psi(\boldsymbol{r},t)$ in each Trotter step, as a function of time for the evolution of a two-dimensional BEC, represented by a Gaussian wave packet $\psi_0(x, y) \propto \exp(-(x^2 + y^2)/2)$, in four different optical trap potentials: (a) an isotropic unmodulated potential with $A_x/W^2 = A_y/W^2 = 0.01$, (b) an anisotropic and anharmonic potential with $A_x/W^2 = 0.01$, $A_y/W^2= 0.015$ and $A_{xy}/W^2 = 0.012$, (c) an anisotropic potential with a four-fold rotationally symmetric modulation ($A_x/W^2 = 0.01$, $A_y/W^2 = 0.015$, $A=10$ and $\boldsymbol{q_1} = (1, 0)$, $\boldsymbol{q_2} = (0, 1)$) and (d) the isotropic eightfold rotationally symmetric potential described in Fig.~\ref{fig:2D_mod}. We use a tensor train of length $40$, modelling a $[-100, 100]^2$ domain with $2^{20}\times 2^{20}$ increments, a non-linearity $g=5$ and time steps $h_t = 0.01$. For spatial evolutions, we impose a maximum bond dimension of $50$ and for the non-linear evolution a tolerance of $10^{-8}$.}
    \label{fig:2D_performance}
\end{figure}

We finally analyze the effects of the non-linearity on the evolution of the bond dimension in two dimensions, which is shown in Fig.~\ref{fig:2D_performance}. Here, we summarize four different optical trap potentials $\hat V(x, y)$: (a) an isotropic unmodulated potential with $A_x/W^2 = A_y/W^2 = 0.01$, (b) an anisotropic and anharmonic potential with $A_x/W^2 = 0.01$, $A_y/W^2= 0.015$ and $A_{xy}/W^2 = 0.012$, (c) an anisotropic potential with a four-fold rotationally symmetric modulation ($A_x/W^2 = 0.01$, $A_y/W^2 = 0.015$, $A=10$ and $\boldsymbol{q_1} = (1, 0)$, $\boldsymbol{q_2} = (0, 1)$) and (d) the isotropic eightfold rotationally symmetric potential described in Fig.~\ref{fig:2D_mod} in the preceding paragraph. Fig.~\ref{fig:2D_performance} shows the maximum bond dimension of the object $U_g \, \psi(\boldsymbol{r},t)$ as a function of the evolution time $T$, which follows from imposing a per-tensor relative error of $\tau = 10^{-8}$. After each complete Trotter step, we truncate the bond dimension of the resulting QTT back to $\chi=50$, which is implemented in the MPO-MPS multiplication of the spatial time-evolution operator $U_V$. Across the four different potentials, we see a clear levelling off of the increased bond dimension due to the non-linear term (for $T \gtrsim 2-3$), with the modulated potentials being more expensive to work with than the unmodulated ones. Here, they typically stabilize in a range between $100$ and $200$. 

\section{Conclusion}
\label{Sec_conclusion}

Here we have demonstrated a methodology to solve the
Gross-Pitaevskii equation, a non-linear partial differential equation, based on tensor networks and the quantics tensor cross interpolation algorithm. Conventional approaches based on numerical finite-difference schemes are limited by the potentially high spatial resolution needed to capture all relevant features of the solution, which implies an exponential memory cost. Our approach, based on the combination of TCI and QTT, circumvents this problem by representing the coarse-grained function as quantics tensor trains, whose number of tensors is the logarithm of the desired spatial resolution. By increasing the bond dimensions at the length scales relevant for the problem, we can achieve an efficient compression of the coarse-grained solution. We have implemented the time evolution by Trotterizing the time evolution operator of the GP Hamiltonian and constructing the tensor network representation of the non-linear operator.

To demonstrate our methodology, we have simulated the dynamics of a Bose-Einstein condensate in various modulated optical trap potentials. We have obtained high-fidelity solutions for potentials ranging over six orders of magnitude by setting a space discretisation of $2^{30}$ increments at a fraction of $10^{-6}$ of the memory required for a conventional vectorized finite-difference approach. Furthermore, we have presented high-resolution solutions of the GP equation in two dimensions, using a grid size of $2^{20} \times 2^{20}$ points. As an example, we have used this to model the dynamics of a Bose-Einstein condensate in a quasicrystalline optical trap, realized in recent experiments. 

Our manuscript establishes a methodology for solving non-linear time-dependent partial differential equations with unprecedented precision at a fundamentally reduced computational and memory cost. Further algorithmic improvements could be achieved by dynamically tuning the bond dimension on the fly, including allowing for a slight increase with time if major deformation of the initial state occurs. Furthermore, as a state in each step of a Trotterized time evolution is evolved to a state with very high overlap to the input state, different tensor train time evolution algorithms, such as TDVP, could be considered as an alternative. Our approach can be readily extended to generic non-linear time-dependent differential equations, providing a strategy to solve generic problems in physics and engineering with unprecedented accuracy by leveraging a quantum-inspired tensor network algorithm. 

\vspace{5mm}

\textbf{Acknowledgments:}
M.N. and J.L. acknowledge financial support from InstituteQ, Jane and Aatos Erkko Foundation, Aalto Science Institute, Research Council of Finland projects No.~331342 and 358088, the Finnish Quantum Flagship, and the ERC Consolidator Grant ULTRATWISTROICS (Grant agreement no. 101170477). We acknowledge the computational resources provided by the Aalto Science-IT project. X.W. acknowledges the funding of Plan France 2030 ANR-22-PETQ-0007 “EPIQ”, the PEPR “EQUBITFLY”,the ANR “DADI”, the ANR TKONDO and the CEA-FZJ French-German project AIDAS for funding. We thank N. Jolly and C. Flindt for useful discussions.

\bibliography{biblio}

\end{document}